\documentclass[12pt]{iopart}
\usepackage{graphicx}
\usepackage{bm}
\usepackage[active]{srcltx}
\usepackage{dcolumn}
\usepackage[usenames]{color}
\usepackage[english]{babel}

\begin{document}
\title[Magnetism in metal-benzene sandwiches and wires]{Interplay of 
structure and spin-orbit strength in magnetism of metal-benzene
sandwiches: from single molecules to infinite wires}
\author{Y~Mokrousov$^{1,2}$,~N~Atodiresei$^2$, 
G~Bihlmayer$^2$, S~Heinze$^1$ and S~Bl\"ugel$^2$} 
\address{$^1$Institute for Applied Physics, University
of Hamburg, Jungiusstrasse 9a, 20355 Hamburg, Germany}
\address{$^2$Institut f\"ur Festk\"orperforschung, 
Forschungszentrum J\"ulich, D-52425 J\"ulich, Germany}
\ead{ymokrous@physnet.uni-hamburg.de}
                                                        
\begin{abstract}
Based on first-principles density functional theory calculations
we explore electronic and magnetic properties of experimentally 
producible sandwiches and infinite wires made of repeating benzene
molecules and transition-metal atoms of V, Nb, and Ta. We describe
the bonding mechanism in the molecules and in particular concentrate
on the origin of magnetism in these structures. We find that all the considered
systems have sizable magnetic moments and ferromagnetic spin-ordering, with
the single exception of the V$_3$Bz$_4$ molecule. By including the spin-orbit
coupling into our calculations we determine the  easy and hard axes
of the magnetic moment, the strength of the uniaxial magnetic anisotropy energy
(MAE), relevant for the thermal stability of magnetic orientation, and the
change of the electronic structure with respect to the direction of the 
magnetic moment, important for spin-transport properties. 
While for the V-based compounds the values of the MAE are only of
the order of 0.05--0.5~meV per metal atom, increasing the spin-orbit strength
by substituting V with heavier Nb and Ta allows to achieve
an increase in anisotropy values by one to two orders of magnitude.
The rigid stability  of magnetism in these compounds together
with the strong ferromagnetic ordering makes them attractive candidates
for spin-polarized transport applications. For a Nb-benzene infinite
wire the occurrence of ballistic anisotropic magnetoresistance 
is demonstrated.
\end{abstract}
\maketitle

\section{Introduction}
\label{1}

Conventional electronics based on charge transport is facing major
challenges as the scale of electronic devices will reach physical
limits where purely quantum phenomena, such as electron spin, may
play a dominant role. Consequently, the new discipline of spintronics has emerged, 
focusing on processes where information is carried by the electron spin
in addition to, or in place of, electron charge. The use of both
charge and spin degrees of freedom is expected to enable revolutionary
technologies due to ultra-high density magnetic recording, low energy
consumption and short access times for reading and writing in
spin-electronic devices~\cite{Maekawa2001:1,Nature1:99,Science278:1997,Science282:1660}.

An important class of  representative magnetic materials which exhibit
novel properties with promising applications in spintronics includes organometallic
molecular magnets. In designing molecular magnets several basic requirements should
be fulfilled: besides a ferromagnetic coupling of local magnetic moments necessary
for coherent transport, a significant thermal stability of magnetism is requested.
A cornerstone quantity which is directly related to the stability of magnetism
against quantum tunneling and thermal fluctuations is the magnetic anisotropy energy (MAE), 
the energy, necessary to change the magnetization direction relative to real space coordinates.
While in molecular magnets with a small number of magnetic ions and significant distances
between them the dipolar contribution to MAE can be safely neglected, spin-orbit coupling (SOC),
which greatly increases in low-coordinated systems, is responsible for the anisotropy barrier.

Much experimental and theoretical effort has been devoted to understanding the spin dynamics
and tunneling rates in organometallic molecular magnets~\cite{Sessoli:1}, however,
surprisingly little attention has been paid so far to the first-principles calculation of their 
electronic and magnetic properties such as magnetic moments, spin 
coupling and magnetic anisotropies, which can be directly related to the measured quantities.
Because of the need for an atomic scale description, {\em ab initio} density functional theory
(DFT) is a driving force for spintronics, and, so far represents the most natural way towards
basic understanding and the design of new spintronics materials. The flexibility and accuracy
in describing elements from the whole periodic table by DFT results in an enormous predictive
power, especially efficient when applied to organometallic compounds, where elements with
properties of a completely different physical origin are involved.

Being still at a very early stage of development, the one-dimensional (1D)
magnetic organometallic sandwiches represent an important class of novel organic molecular
magnets, which have been suggested as ideal candidates for the smallest possible high-spin
magnets, allowing an extremely high density data storage. Several experimental and theoretical
studies have been performed on magnetic sandwiches formed by vanadium and benzene, V$_{n}$Bz$_{m}$
(see Refs.~\cite{Wang2005:7, Muetterties1982:14,Weis1997:12, Bauschlicher1992:11, Pandey2000:13,Pandey2001:13.05}
and citations therein). The EPR measurements on VBz (half-sandwich) and VBz$_{2}$ estimate magnetic
moments of $1\mu_{B}$~\cite{Mattar1997:7.1}. Stern-Gerlach type magnetic deflection experiments on
V$_{n}$Bz$_{n+1}~(n<5)$ complexes suggest an increase of magnetic moment with the number of V atoms
in the molecule~\cite{Miyajima2004:7.15}, indicating that the magnetic moments of the V atoms couple
ferromagnetically. Synthesis, infrared spectroscopy~\cite{Lyon2005:7.2} and electric dipole
measurements~\cite{Rabilloud:03} on MBz (half-sandwich) and MBz$_{2}$~(M = Nb,~Ta) compounds have
also been reported.  Although these experiments have been carried out on molecules in the gas
phase, magnetically similar and structurally stable molecules suitable for the use of deposition
techniques on substrates in ultrahigh vacuum atmosphere can be synthesized. A possible example are
the $\sigma,\pi$-Donor/$\pi$-Acceptor metal complexes such as bis($\eta^6$-arene) and
bis[$\mu$-($\eta^6$:$\eta^6$-biphenyl)] molecules which can only be prepared by means of cocondensation
techniques~\cite{Elschenbroich}. Concerning theoretical investigations of the magnetism in
V$_{n}$Bz$_{n+1}$ clusters several applications in spintronics of these molecules have been recently 
suggested~\cite{Maslyuk:06,Brandbyge:07}.

In this paper, based on our {\em ab initio} full-potential DFT calculations,
we present a complete and detailed picture of the electronic and magnetic
properties of the M$_{n}$Bz$_{m}$~(M =~V,~Nb,~Ta) sandwiches and infinite
wires~\cite{Mokrousov:IAP,Mokrousov:IJQC}, obtained by an infinite 
repetition of the metal-benzene half-sandwiches. In particular, we concentrate
on the values of the magnetic anisotropy energy barrier, a 
value of an utter importance for technological applications. While most 
of the V-based complexes have a strong ferromagnetic ordering of the
spins, they reveal quite small values of MAE. Substituting vanadium ions 
with the heavier magnetic Nb and Ta ones allows to tune the spin-orbit strength with consequent
values of MAE two orders of magnitude larger, preserving strong ferromagnetic coupling of the spins.
Moreover, due to the position of the Fermi level and significant spin-orbit 
splitting of the bands, infinite NbBz wires are suggested to exhibit the recently
proposed~\cite{Velev2005:7.4} ballistic anisotropic magneto-resistance (BAMR) effect,
i.e.\ the change of the ballistic conductance upon changing the magnetization direction,
which opens the vista for new device concepts in spintronics.

Most of the compounds considered in this paper have
a magnetization which lies in the plane of the benzenes,
perpendicular to the symmetry axis of the molecules ($z$)
with negligible total energy variations as a function of
spin-rotation around the $z$ axis. This fact will
make it difficult to
stabilize the magnetism in these complexes against thermal
fluctuations despite the fact that the energy needed to
switch the magnetization direction from in-plane to the
$z$-direction can be rather large. However, for the
V$_{3}$Bz$_{4}$, V$_{4}$Bz$_{5}$ molecules and a Nb-Bz
infinite wire, the magnetization direction points along the
$z$-direction and the values of the uniaxial anisotropy
are sizable. This suggests that a large subclass of the
M$_{n}$Bz$_{m}$ ($n,m=1,2,...$) complexes has a magnetization
which is bi-modal in nature and may be applicable,
e.g., for data storage or as qubits for quantum information,
owing to the rigid fixation of the spin
moments in these compounds along the single axis in space.
The latter properties in combination with the small size,
transparent electronic structure and large magnetic moments
make this subclass valuable for potential applications in future spin-electronics.

\section{Method and computational details}
\label{2}
The calculations were performed within the framework of
density functional theory (DFT) using the generalized
gradient approximation (GGA-PBE)~\cite{Perdew:7.5} for the
exchange-correlation potential. We employed a realization
of the full-potential linearized augmented plane-wave method
for truly one-dimensional systems (1D-FLAPW)~\cite{Mokrousov2005:4},
as implemented in the \texttt{FLEUR} code~\cite{FLEUR:5}. All
the isolated molecules were calculated using unit cells repeated along
the symmetry axis ($z$-axis) with a large spacing between molecules in order
to exclude unwanted side effects on the electronic and magnetic
structure due to interaction with the neighbors. The smallest
distance between the closest atoms in two neighboring molecules was
larger than 13.5~a.u.~in our calculations.
The $z$-axis of the system cuts through the transition-metal atoms and the
centers of gravity of the Bz molecules.
The following muffin-tin radii were chosen: 2.4~a.u.~(V,~Nb,~Ta),
1.25~a.u.~(C) and 0.65~a.u.~(H).
In the vacuum the wave functions were expanded into cylindrical 
basis functions with an angular expansion up to $m_{\rm{max}}=24$.
Correspondingly, the charge density and potential were expanded up to
$m_{\rm{max}}=48$.
For calculating the isolated molecules we used one $k$-point and for the 
infinite wires we used 8 and 16 $k$-points in one half of the Brillouin zone,
respectively. The structures were relaxed with the basis
functions cut-off parameter $k_{\rm{max}}$ of 3.5~a.u.$^{-1}$,
and for calculations of the magnetic ground-state at the relaxed
positions converged values of the spin moments and total
energies were achieved with $k_{\rm{max}}$=3.6~a.u.$^{-1}$. 
As matter of convenience, throughout the paper we use the expression
'Fermi energy' not only to indicate the highest occupied state of an
infinite metallic wire, but also to indicate the highest occupied orbital
in single molecules, keeping in mind that for isolated molecules it
looses the meaning of a chemical potential. 

The magnetic anisotropy-induced energy barrier which needs to be overcome to change
the magnetization direction from one which is lowest in energy (easy axis) to the
energetically most unfavorable one (hard axis) is called the magnetic anisotropy
energy (MAE). MAE consists of the two contributions: due to classical dipole-dipole
interaction of local magnetic moments and due to the spin-orbit coupling.
For molecules with few magnetic centers and magnetic wires the dipolar
interaction is very small, and the role of spin-orbit coupling becomes crucial, 
leading to huge values of orbital moments and
magnetic anisotropy energy~\cite{Mokrousov2006:4}.
By including spin-orbit coupling in the total energy
calculations we considered two possible symmetry-determined
directions of the magnetization in the molecules:
along the $z$-axis ($z$-case) and in radial direction parallel to the planes of
the benzenes ($r$-case), with the most energetically preferable
axis being the easy axis. MAE
is defined as a total energy difference between these two configurations.
For the molecules and wires considered in this paper, the dipolar contribution
to the MAE amounts to not more than a tiny value of 0.01~meV per metal atom, with
the $z$-axis as an easy axis. Therefore, we restrict our discussion in the following to the SOC-part
of the MAE.
The variation of the MAE as a function of the $r$-axis in the plane
of the benzenes normally contributes only a small fraction of the MAE value
and is neglected.
The magnetic anisotropy energy
was obtained using the force theorem with the number
of $k$-points up to 64, and the values of $k_{\rm{max}}$
up to 4.0~a.u.$^{-1}$, reaching convergence in
the calculated values.  

\section{General remarks}
\label{3}
\begin{figure*}[ht!]
\begin{center}
    \includegraphics[scale=0.55]{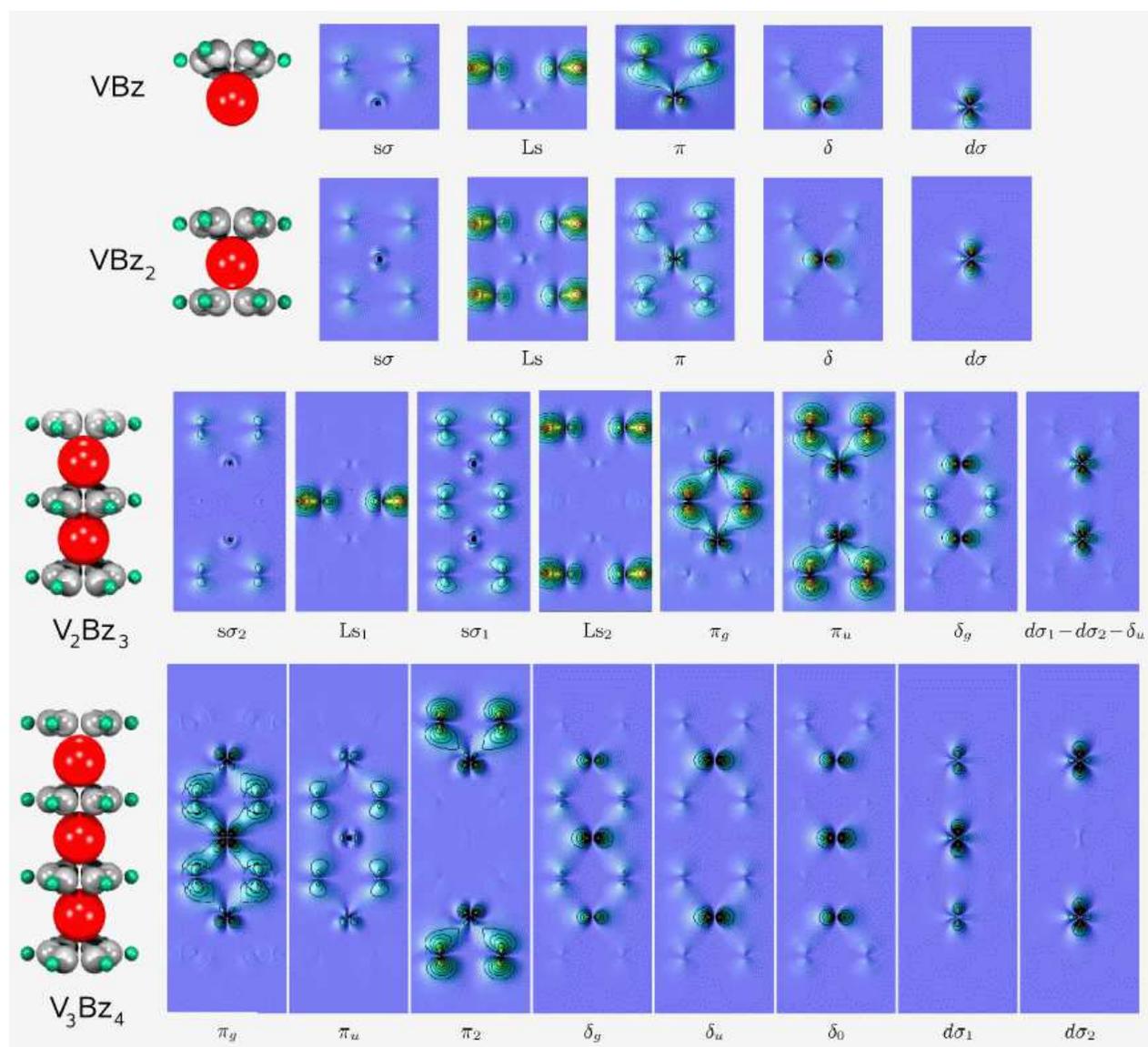}
\end{center}
\caption{Geometrical structure of the V-Bz complexes (V-red,
C-gray, H-green) and the charge density plots of the
orbitals from figures~2,~3,~6 in the plane cutting through V,
C and H atoms ($x-z$ plane).}
   \label{fig3}
\end{figure*}

The geometries of the considered molecules V$_{n}$Bz$_{n+1}$ with $n=1,2,3$ are
shown in figure~\ref{fig3}.
Our calculations show that stable
structures of the molecules are sandwiches with the metal
atom above the center of gravity of Bz molecules, which
was also found in other theoretical and experimental
investigations (see Ref.~\cite{Wang2005:7, Miyajima2004:7.15, Miyajima2005:7.16}
and citations therein). The geometry of the metal-benzene
molecules was restricted to $C_{6v}$ symmetry, while
all other structural parameters were relaxed.
In general, spin-polarization does not change the nonmagnetic
equilibrium structural values by more than 1-2$\%$. For
all sandwiches the planarity of the outer Bz molecules is
broken with the planes of C and H hexagons parallel but
slightly shifted relative to each other along the $z$-axis (the value
of the shift ranges between 0.02 and 0.09~a.u., depending on the compound).
For the V$_2$Bz$_3$ molecule we analyzed the changes in the electronic
and magnetic structure due to rotations of the benzene
molecules with respect to each other around the $z$-axis.

Structural optimization of the single benzene molecule yields the values of
2.625~a.u.~for the C-C bond length and 2.058~a.u.~for the C-H bond length,
in perfect agreement with the values obtained with various methods and
reported in Ref.~\cite{Bauschlicher1992:11}.
While the changes in the C-H bond length due to the interaction with the 
transition-metal (TM) ions are negligible, the C-C distance in the 
benzene rings increases by at most 4.3\% for the Ta infinite wire.
As a function of the transition metal, the C-C bond length increases
from 2.3\% to 4.1\% for half-sandwiches and from 3.4\% to 4.3\% for the 
infinite wires, going from V to Ta. Concerning the distance between the
metal atoms and the centers of the closest carbon rings (M-C distance) 
in V-based molecules,
significant increase from the value of 3.075~a.u.~for the half-sandwich
up to 3.622~a.u.~(19\%) in the V$_3$Bz$_4$ molecule is observed, revealing
a strong inhomogeneity as a function of the relative position of the 
V atom with respect to the center of the molecule. 
Changing the atomic number of the metal ion in the half-sandwiches
and infinite wires results in the increase in the M-C distance
by at most 2\%, as compared to V corresponding values.

In a simple picture the mechanism responsible for bonding
in the molecules can be given on the basis of a schematic
analysis of the orbitals of the metal atom (V,~Nb,~Ta) and benzene molecule,
which can be classified in terms of the pseudo-angular momenta
around the symmetry $z$-axis. Five $d$ orbitals of the metal atom
can be divided according to their symmetries into one
$d\sigma\,(d_{z^2})$, two $d\pi\,(d_{xz},d_{yz})$, two
$d\delta\,(d_{x^2-y^2},d_{xy})$. 
Six $\pi$-orbitals of Bz 
form one $L\pi\,(\pi_1)$, two degenerate $L\pi\,(\pi_2,\pi_3)$
(HOMO), two degenerate $L\delta\,(\pi_4^{*},\pi_5^{*})$ (LUMO)
and one $L\phi\,(\pi_6^{*})$ orbital. When Bz molecules
and metal atoms are brought together, the  HOMO and LUMO orbitals
of Bz interact with the metal $s,d$ orbitals of the same
symmetry and bonding occurs~\cite{Fleming1976:8} (see also figure~\ref{fig3}).
Yasuike~\cite{Yasuike1999:9} explained the electronic structure
of the V$_{n}$Bz$_{n+1}$ complexes based on a schematic
orbital interaction diagram using extended H\"uckel or
Hartree-Fock methods. We performed the DFT electronic 
structure calculations for these complexes with a side goal of
checking the validity and limitations of the latter
computational models. 

In case of spin-polarized calculations, the calculated spin moments
inside the muffin-tin spheres of carbon and hydrogen atoms constitute
almost constant values of 0.01$\mu_B$ and 0.001$\mu_B$, respectively.
Therefore, further on we specify only the direction of the carbon 
local spin moments in considered compounds.

Despite a great advantage of the DFT in predicting the trends in the
properties of a large set of V$_{n}$Bz$_{n+1}$ complexes, we are well
aware of the fact that in low-coordinated systems localized states develop,
for which the common approximation to the exchange-correlation
functional, the local density approximation (LDA) or the GGA, 
does not always provide reliable enough results. 
On the other hand, molecules used in context of spintronics or molecular
electronics are typically attached to metallic leads. Electrons in the leads
contribute by screening to the reduction of the Coulomb interaction in a molecule 
and the actual strength of the Coulomb interaction in this situation remains 
a subject of further investigations.
The self-interaction correction of a singly-occupied $d\sigma$ state and the
Coulomb interaction of localized states in molecules attached to the leads
may be taken care of in the future by more reliable functionals, which
came currently under
scrutiny.

\section{$\rm{V-Bz}$ molecules}
\label{4}
We start with the study of the electronic and magnetic properties
of the V$_{n}$Bz$_{n+1}$ sandwiches, paying special attention
to the origin of magnetism in these compounds and its development
with increasing number of V atoms in the complexes. The issue of 
stability is addressed and a possibility of experimental observation of 
magnetism in these compounds
is discussed via evaluating the values of magnetic anisotropy. 
The insight into the physical properties of the V-Bz molecules which we provide
in this section is important for understanding the trend in magnetic   
and electronic properties of the compounds with Nb and Ta ions,
considered in the next section.
\subsection{Half-sandwich VBz}
\label{4.1}

\begin{figure*}
\begin{center}
    \includegraphics[scale=0.67]{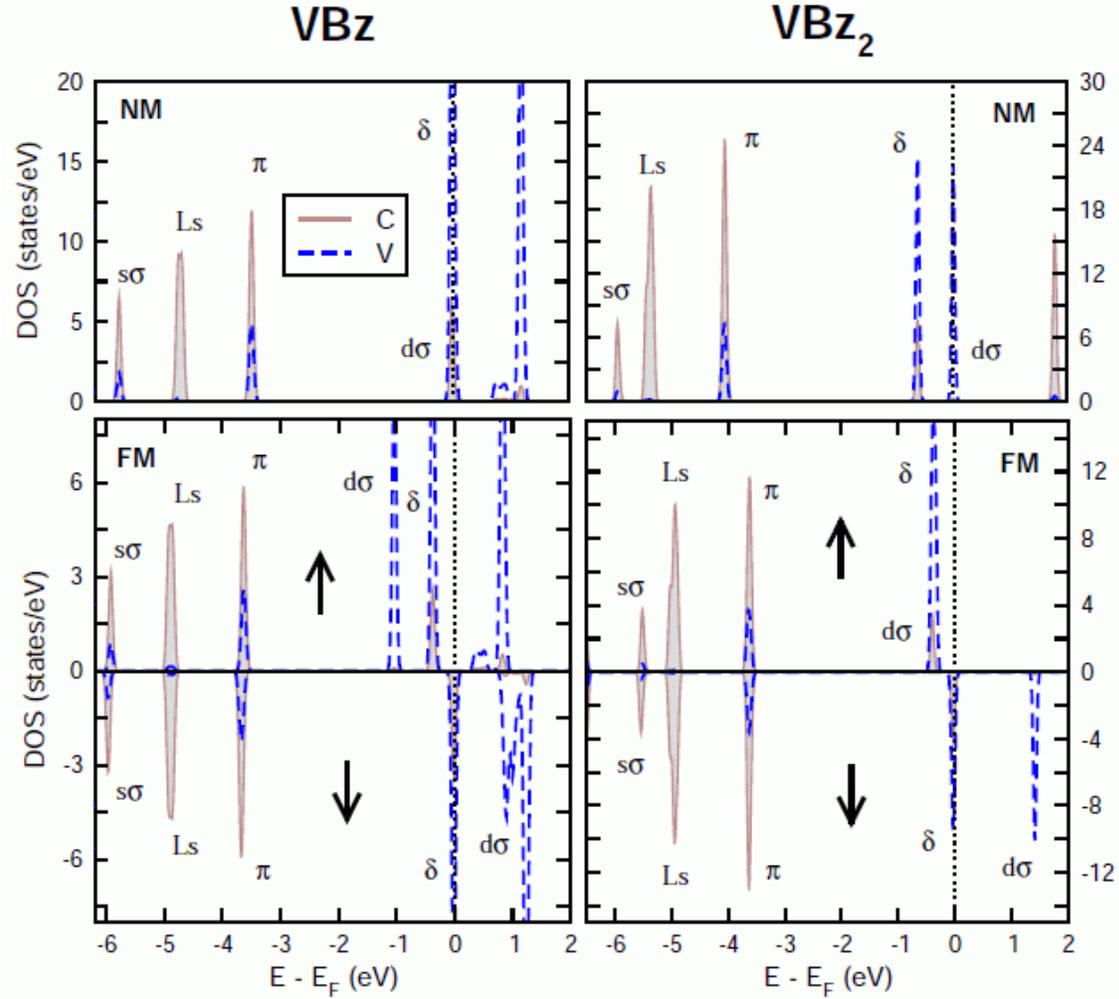}
\end{center}
   \caption{Local density of states (LDOS) for the VBz and VBz$_2$ molecules. Spin up and spin down channels for the
 spin-polarized case are indicated by arrows. 
 The charge density plots of the states for the NM case are presented
in figure~\ref{fig3}.}
   \label{fig1}
\end{figure*}

In order to understand the bonding and formation of a magnetic moment 
in the VBz molecule, we take a look at the local
densities of states (LDOS) for the non-magnetic (NM) and
spin-polarized (FM) cases at the optimized atomic positions, 
presented in figure~\ref{fig1} (left panel).
In the NM case the nonbonding $d\sigma$ orbital
at the V site is situated at the Fermi level and accommodates
one electron. The interaction between  the LUMO orbitals
of Bz and the $d\delta$ orbitals of the V atom results in
two degenerate bonding $\delta$ states slightly lower in
energy  which accommodate four electrons located mostly at
the V site. The overlap of the $d\pi$ orbitals of the V atom
and $L\pi(\pi_2,\pi_3)$ HOMO orbitals of Bz produces the next
low-lying $\pi$-type states of the VBz molecule
carrying four electrons located mostly at the
C sites of the Bz ring.
The interaction of
the lowest $L\pi(\pi_1)$ orbital of Bz with the $4s$ orbital
of vanadium results in a low lying $s\sigma$ state
accommodating two electrons located mostly at the C sites
of the Bz ring. 
The closest to the HOMO orbitals, low-lying $Ls$ benzene orbital 
is not participating in the interaction with the metal ion.
The charge density plots
for the resulting states of the molecule are shown in figure~\ref{fig3},
which illustrate the orbital character and bonding of the states.

The magnetic solution for this compound is lower in total energy 
than the nonmagnetic one by a large value of 0.491~eV.
The spin-polarized LDOS displays a large exchange splitting
of the states near the Fermi level (figure~\ref{fig1} (left panel)).
This splitting is largest for the $d\sigma$ state of V
(located at the Fermi level in the nonmagnetic DOS). For
the $\delta$ states, involved in the bonding, the exchange splitting
is much smaller. Out of five electrons of V, three occupy 
spin up (one at the $d\sigma$ level and two at the two
degenerate $\delta$ levels) and two occupy spin down
(two degenerate $\delta$ levels) states. This
results in a total spin moment of 1$\mu_B$ for
the system. Our calculations show, that this spin
moment is entirely located inside the muffin-tin sphere
of the V atom (1.09$\mu_B$). There is no significant exchange 
splitting of the low bonding states of the VBz molecule.
The spin moments of the C atoms couple antiferromagnetically 
to the spin moment of the V-atom.

\subsection{Single sandwich VBz$_2$}
\label{4.2}

When two Bz molecules are placed at the same distance from
V, as in the VBz$_2$ molecule, the symmetry-adapted molecular
$L(s,\pi,\delta,\phi)_{g,u}$ orbitals of the molecular
complex Bz$_2$ are produced from the orbitals of two Bz molecules.
The atomic orbitals of V have $g$ symmetry and can interact only with
the $L(s,\pi,\delta)_g$ orbitals of Bz$_2$.
According to the H\"uckel interaction scheme
a nonbonding $d\sigma$ orbital occupies the Fermi energy with
one electron. As a results of hybridization between the $\delta$
orbitals of V and the $(L\delta)_g$ orbitals of Bz$_2$ two degenerate
$\delta$ states lower in energy are created and accommodate four
electrons. The overlap between the $\pi$ orbitals of V and $L\pi_g$
orbitals of Bz$_2$ produces two $\pi_g$ bonding orbitals, while
$L\pi_u$ ($\pi_u$) remain nonbonding (for the VBz$_2$ sandwich the
$\pi_g$-$\pi_u$ splitting is barely visible due to a large distance
between the V atom and benzene molecules, as compared to the compounds
with larger number of vanadii, discussed below).
The basic features of the nonmagnetic
LDOS (figure~\ref{fig1} (right panel) with corresponding plots of the 
orbitals in figure~\ref{fig3}) reflect the H\"uckel
type interaction scheme for this compound~\cite{Yasuike1999:9}. 
As in the case of the half-sandwich VBz,
the $d\sigma$ orbital with one electron is located at the
Fermi level. Compared to the VBz LDOS, the double
degenerate  $\delta$  orbitals of the VBz$_2$
molecule are situated much lower in energy. This suggests a
different bonding of the  $d\delta$ vanadium orbitals and
$(L\delta)_g$ orbitals of the molecular complex Bz$_2$.

The magnetic solution for this molecule is lower in total energy than the nonmagnetic
one, with a difference in total energies $E_{NM}-E_{FM}$ of
0.459~eV. The spin-polarized LDOS  of the VBz$_2$ molecule
(figure~\ref{fig1}) reflects the main features of the VBz
DOS. The largest exchange splitting occurs for the $d\sigma$ state,
while the exchange splitting for the $\delta$ states
is significantly quenched. 
The low-lying $\pi_g,\pi_u,s\sigma,Ls$ 
states of the molecule are almost not exchange split.
The total spin moment of the VBz$_2$ molecule constitutes 
the same value of 1$\mu_B$, as in the case of VBz.
Furthermore, the spin moments inside the muffin-tin spheres of V,
C and H  are almost of the same values, as in the case of the
VBz molecule, with antiferromagnetic coupling of the V and C
local moments. Our value for the total magnetic moment agrees
with the experimental average value of 0.8$\mu_B$, measured by Kaya
and co-workers (Ref.~\cite{Miyajima2004:7.15}) for the
VBz$_2$ molecule at room temperature in the
gas phase.

\subsection{Double sandwich V$_2$Bz$_3$}
\label{4.3}

When two outer Bz molecules are at the same distance from an 
inner Bz ring as in the V$_2$Bz$_3$ compound, symmetry-adapted
molecular $L(s,\pi,\delta,\phi)_{g,u}$ orbitals
of the molecular complex Bz$_2$ are created from the orbitals
of two benzenes. The atomic orbitals of the V atoms and the molecular 
orbitals $L(s,\pi,\delta,\phi)$ of the central Bz have
$g$-symmetry. Bonding molecular orbitals of V$_2$Bz$_3$ between
V atoms and Bz molecules are formed due to interactions of the 
orbitals with the same symmetry (e.g. $g$-orbitals). In general,
the orbital interaction scheme for this complex becomes rather 
cumbersome.

\begin{figure*}
\begin{center}
      \includegraphics[scale=0.67]{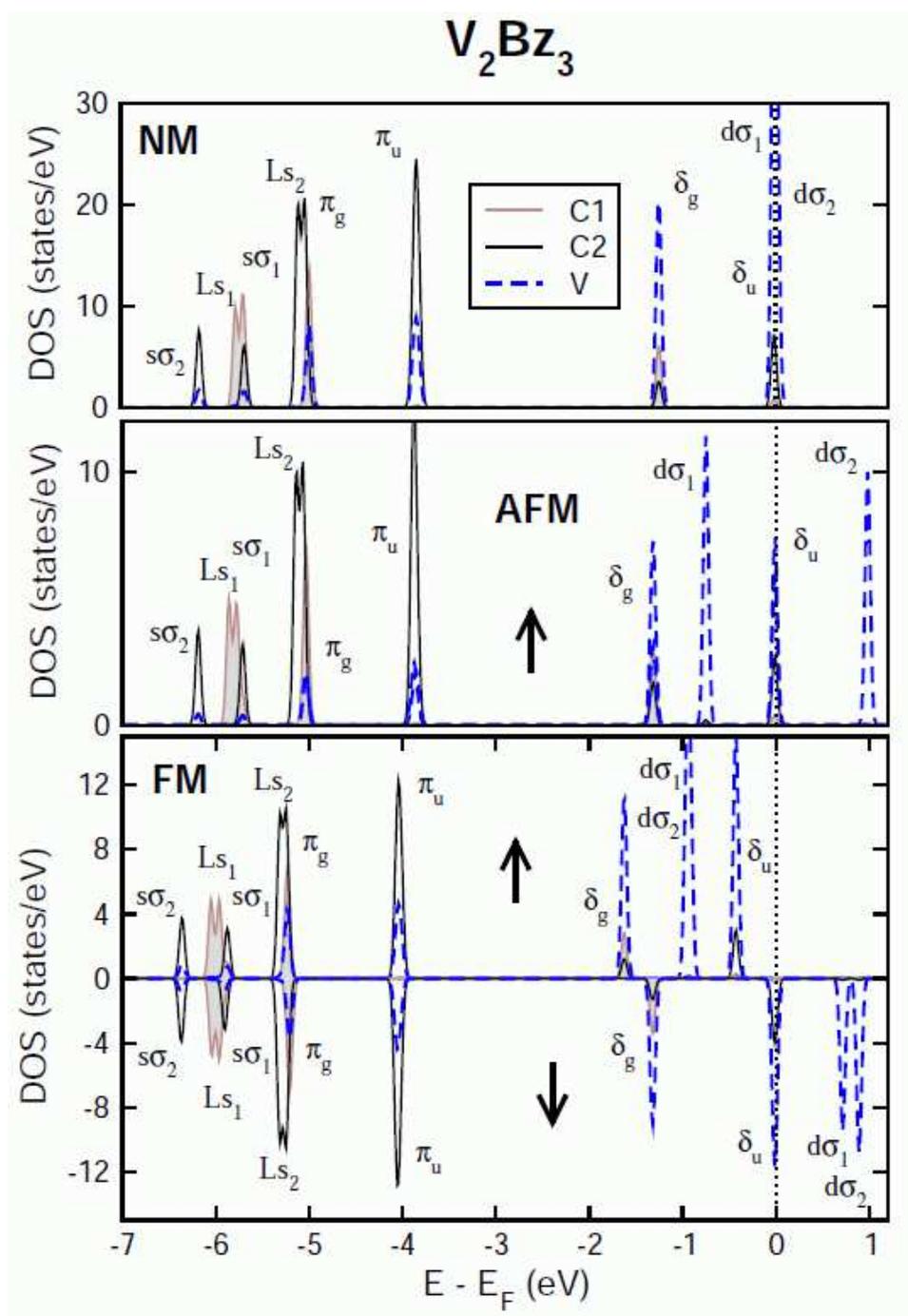}
\end{center}
   \caption{LDOS for the V$_2$Bz$_3$ molecule. Spin up and spin down for the
 spin-polarized case are indicated by arrows. The charge density plots of the states 
 for the NM case are presented
 in figure~\ref{fig3}.}
   \label{fig2}
\end{figure*}

The LDOS of this molecule are presented in figure~\ref{fig2} and 
corresponding charge density plots of the orbitals are shown
in figure~\ref{fig3}. The notations for the orbitals
follow the same scheme as for VBz and VBz$_2$. As a
general observation, orbitals localized at the outer Bz rings
(carrying $u$-index) are situated at higher energies than the
corresponding orbitals of the inner Bz ring (carrying $g$-index).
In the NM case nonbonding $d\sigma_1$ and $d\sigma_2$ orbitals are 
positioned at the Fermi energy, and slightly lower in energy there are 
two degenerate $\delta_u$ orbitals arising from the hybridization
between the $d\delta$'s of V atoms and LUMO of the outer Bz's.
Much lower in energy are two degenerate $\delta_g$ 
orbitals reflecting strong hybridization between $d\delta$'s of V
atoms and LUMO of the inner Bz. Two $\pi$-hybrids, $\pi_u$
and $\pi_g$, are situated very deep in energy followed
by a set of $Ls$- and $s\sigma$-type orbitals. 

Our calculations show, that the V$_2$Bz$_3$ molecule is magnetic,
with a total energy difference between the FM and NM
solutions (to which we refer below as $E_{M}$) 
of 0.415~eV per metal atom. The AFM solution with
opposite vanadium spins is by a tiny value
of 3.0~meV per V atom higher in energy than the FM solution,
suggesting  that the magnetic order of the molecule can be
easily influenced by effects such as thermal fluctuations and
vibrations, above a threshold temperature $T_0$~\cite{Pandey2004:13.1}. 
The total spin moment of the molecule of 0$\mu_B$ (AFM)
and 2$\mu_B$ (FM) for these magnetic states should,
in principal, result in an average value of 1.0$\mu_B$ for
temperatures much larger than $T_0$.
This fact can explain an average value of
1.3$\mu_B$ for the complex, measured at $T=154$K by Kaya and
co-workers~\cite{Miyajima2004:7.15}.

The spin moment value of 1.02$\mu_B$ inside the V muffin-tin
sphere in both FM and AFM configurations is close  
to that of the VBz half-sandwich. In the FM case the spin    
moments of the C atoms are parallel to each other
and opposite to the V spin moments. In the AFM case, the C spin moments
in the outer Bz's  are antiferromagnetically coupled to the closest
V atom, while the C moment in the central Bz vanishes.
Some authors~\cite{Yasuike1999:9} suggest that the induced
carbon spin moment and its orientation with respect to
the moments of vanadii could play a crucial role for the spin
ordering of the latter ones. While the influence of the carbon spin
moments on the magnetic ordering in the molecule is probably negligible,
we performed a calculation, where we removed the central benzene
from the complex, keeping the coordinates of all the other 
constituents constant. In this case, the total energy difference between the
FM and AFM solutions $E_{FM}-E_{AFM}$ (to which we refer below         
as $E_{AF}$) amounts to 22~meV per metal atom in favor of ferromagnetism.
Thus the direct exchange
between the V atoms favors ferromagnetism and the competition
with antiferromagnetic super-exchange via the central benzene
ring results in a small total $E_{AF}$ value of 3.0~meV per metal atom
for the V$_2$Bz$_3$ sandwich.

The FM and AFM LDOS (figure~\ref{fig2}) display large
exchange splitting of the non-bonding V $d\sigma$ states:
in both cases it reaches 2.5~eV. In the FM case
the strongly bonding $\delta_{g}$ orbitals
show smaller exchange splitting than the $\delta_{u}$ states,
while the bonding low-energy $\pi$, $Ls$ and $s\sigma$ orbitals
are not exchange split at all.

\begin{figure}
\begin{center}
 \includegraphics[scale=0.65]{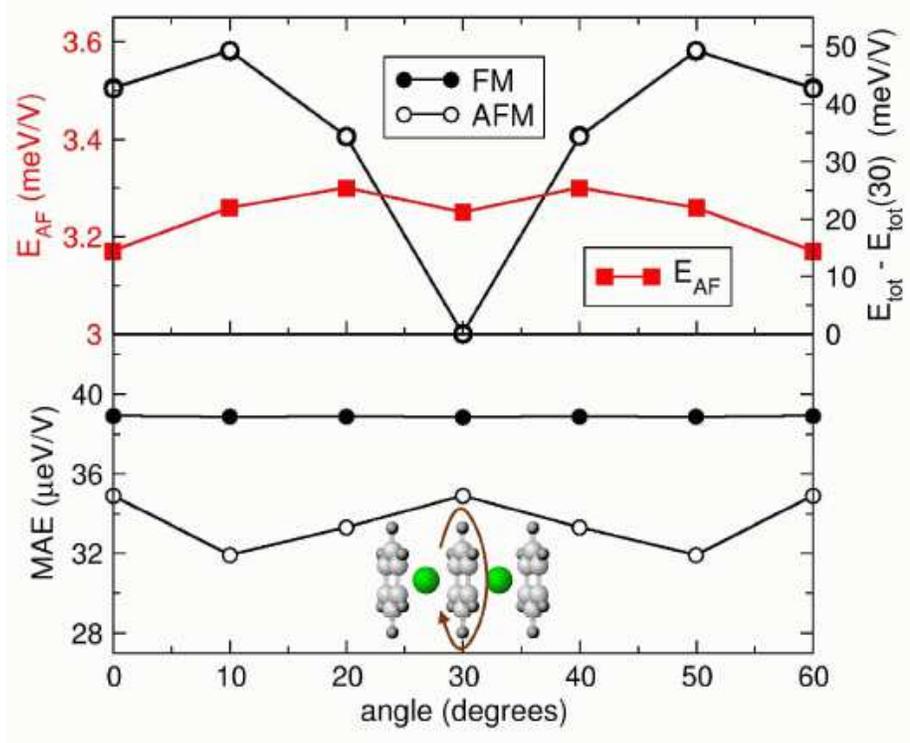}
\end{center}
\caption{Properties of the V$_2$Bz$_3$ molecule as a function
of the angle between the central and outer benzenes.
All values are given per V atom.
Note, that in the upper graph two energy scales are given and variation
of the total energy $E_{\rm{tot}}-E_{\rm{tot}}(30)$ is the same
for the FM and AFM solutions. As opposed to the total energy difference,
variations in the $E_{AF}$ energy and MAE with respect to the angle
between the benzenes are very small.}
 \label{fig4}
\end{figure}

\begin{figure}
\begin{center}
 \includegraphics[scale=0.63]{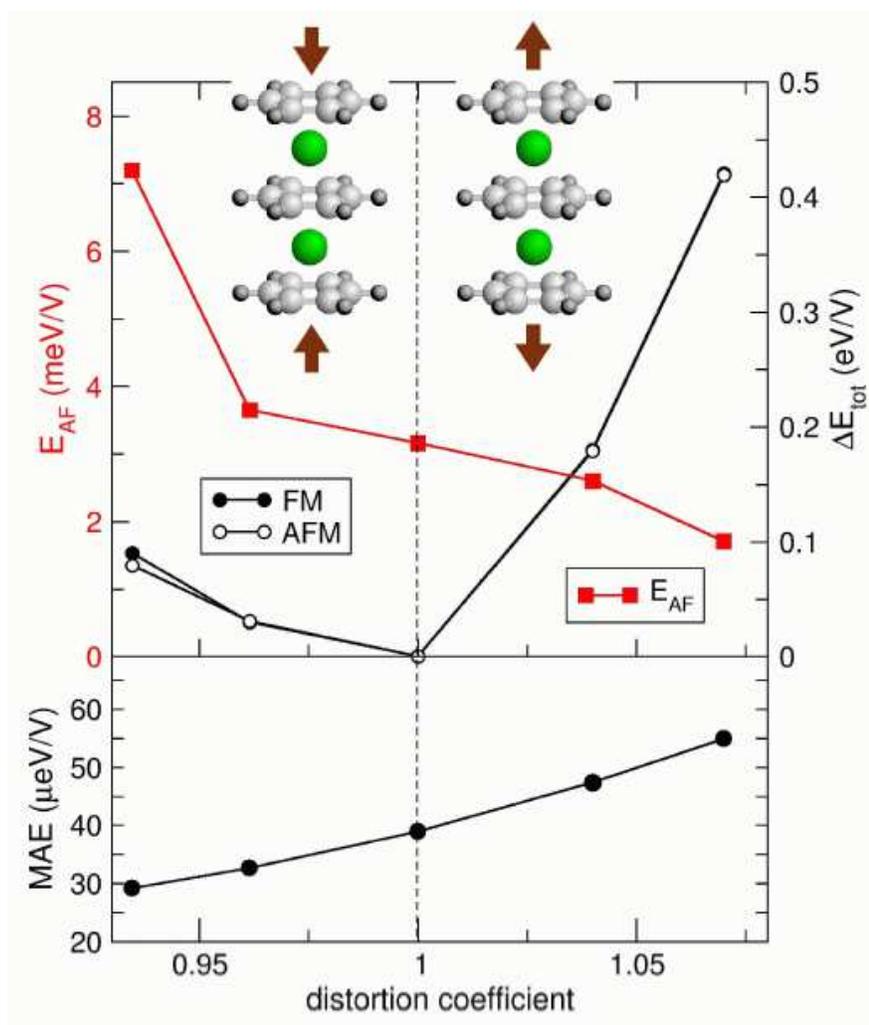}
\end{center}
\caption{Properties of the V$_2$Bz$_3$ molecule as a function 
of the scaling coefficient along the $z$-direction. All
values are given per V atom. As a function of the $z$-scaling,
large variations in the $E_{AF}$ and MAE can be seen. For contracting
molecule direct exchange between the V atoms is responsible for an
increase in the $E_{AF}$ energy. For expanding molecule, the super-exchange
between vanadii via the central benzene causes a significant shift in the
$E_{AF}$ energy towards antiferromagnetism.}
   \label{fig5}
 \end{figure}

In order to understand the influence of the molecules conformation on the
spin coupling in this sandwich we investigated the changes in the electronic
and magnetic properties with respect to the rotation of the central
benzene ring (figure~\ref{fig4}) and $z$-distortion of the sandwich (figure~\ref{fig5}). 

In the first case we rotated only the atoms of the central benzene
ring by an angle $\theta$ around the $z$-axis and fixed all other
atom positions. The total energy $E_{\rm{tot}}(\theta)$, figure~\ref{fig4},
shows a minimum at an angle of $\theta=30^\circ$ for both the FM and
the AFM state. The significant energy gain of 50 meV suggests
that the actual ground state might resemble a non-trivial angular
orientation of the benzenes in the molecules (indeed, the chirality
of the complexes was reported in Ref.~\cite{Wang2005:7}). Quite surprisingly the energy
difference $E_{AF}$ between the FM and AFM solution, which is one order of
magnitude smaller than the total energy variation, does not display any
significant changes. The same is true for the value of the magnetic
anisotropy energy (MAE), which is in many systems extremely sensitive to the
slightest deviations in the electronic structure.

In order to model contractions and elongations of the sandwiches we fixed
all in-plane coordinates while the $z$-coordinate of all atoms was multiplied
by a factor $\alpha$ (the central benzene ring acts as a reference point).
Such contractions and elongations along the $z$-axis result in large deviations
of the total energies for the FM and AFM solution, the energy difference
$E_{AF}$, and the values of the MAE, as shown in figure~\ref{fig5}. The total
energy has its minimum in the unstretched case ($\alpha=1$) and rises by large
values of 50 meV upon contraction ($\alpha=0.95$) and 250 meV upon
elongation ($\alpha=1.05$). The magnetic coupling is also strongly
influenced. For an elongated sandwich ($\alpha>1$) the antiferromagnetic
super-exchange between the V atoms via the central benzene ring becomes
even more pronounced and the energy difference $E_{AF}$ drops by almost
half its value. Accordingly, the direct exchange between the V atoms dominates for a
contracted molecule ($\alpha<1.0$) and $E_{AF}$ increases by 4~meV in favor of
ferromagnetism.

\subsection{Triple sandwich V$_3$Bz$_4$ and V$_4$Bz$_5$}
\label{4.4}

For the triple sandwich V$_3$Bz$_4$ we considered three possible magnetic
configurations: $\uparrow\uparrow\uparrow$--ferromagnetic (FM),
$\uparrow\downarrow\uparrow$--antiferromagnetic (AFM1), and
$\uparrow\uparrow\downarrow$--antiferromagnetic (AFM2). 
The FM-state is by 300~meV per V atom lower in energy
than the nonmagnetic solution. Among the magnetic solutions,
the total energy difference between the FM and AFM1 configurations
of 4.1~meV per V atom favors the $\uparrow\downarrow\uparrow$-state, while the AFM2 $\uparrow\uparrow\downarrow$-state
is by 1~meV (per V) higher in energy than the FM state.
However, the energy barrier of 12.3~meV for the whole complex
can be easily overcome due to thermal fluctuations, increasing
the total spin moment of the system by occupying the excited high-spin
FM state. This leads to an increase of the average value 
of the spin moment with increasing temperature which was indeed observed
experimentally~\cite{Miyajima2005:7.16}. 

In the FM case, the spin moments of the central and outer vanadii
constitute 0.76$\mu_B$ and 1.17$\mu_B$, respectively, resulting
in a total spin moment for the whole molecule of 3$\mu_B$. The transfer
of the spin density between V atoms indicates the delocalization
of the $d_{z^2}$-electrons along the $z$-axis. The magnetic moments
of the carbon atoms are antiparallel to the vanadium moments.
For the $\uparrow\downarrow\uparrow$-configuration, 
the total spin moment of the system 
constitutes 1$\mu_B$, with an absolute value of 1.07$\mu_B$ inside
every V sphere, indicating a higher degree of electron localization on V sites. 
The moments of the outer carbons are opposite to  the closest V spins,
while the carbon moments in the two central rings of the complex vanish.  
 
\begin{figure}
\begin{center}
      \includegraphics[scale=0.64]{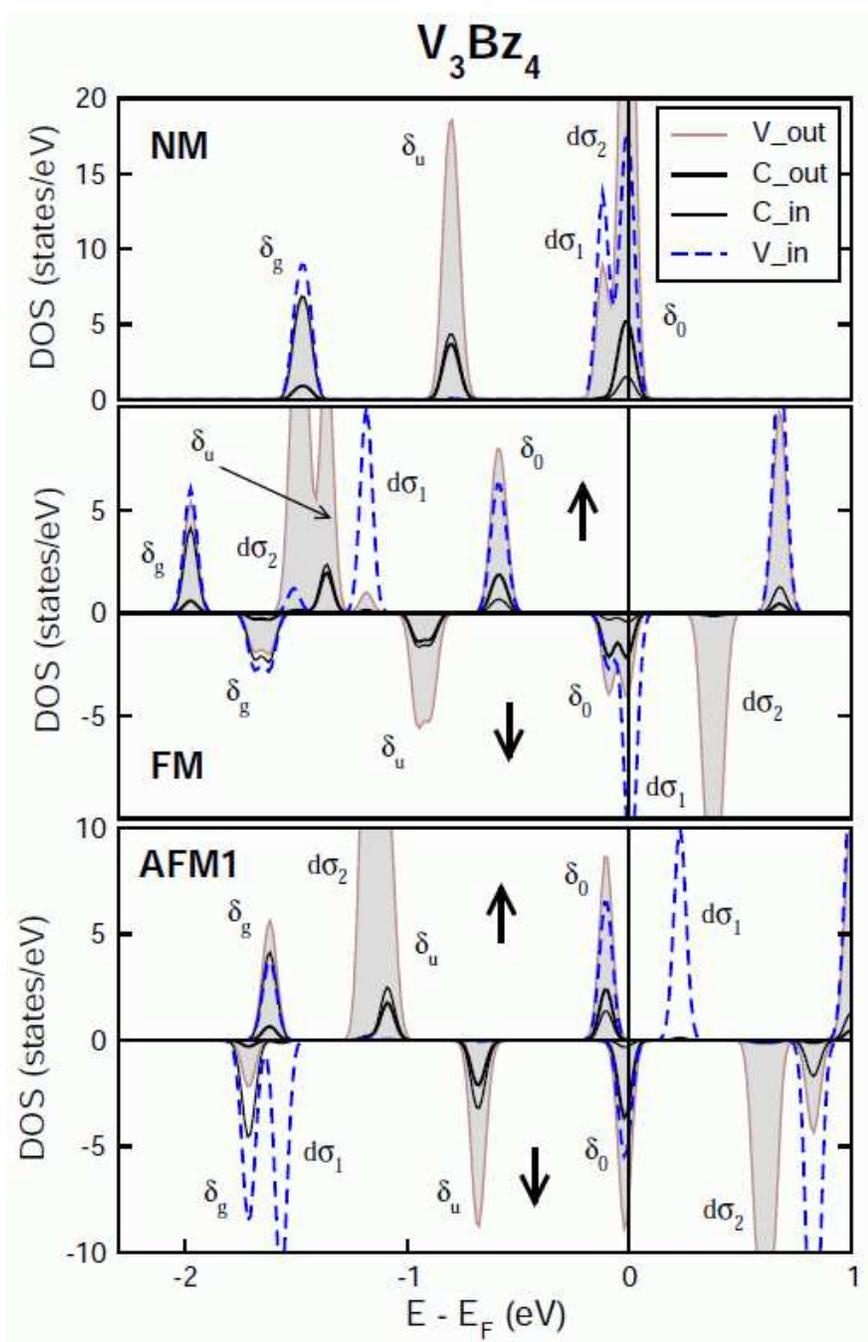}
\end{center}
  \caption{LDOS for the V$_3$Bz$_4$ molecule in the NM, FM and AFM1
configurations. Spin-up and spin-down are indicated by arrows.
The charge density plots of the states for the NM case are presented 
in figure~\ref{fig3}.}
   \label{fig6}
 \end{figure}

The LDOS for the NM, FM and AFM1 solutions are presented in 
figure~\ref{fig6}  
and the charge density plots of the marked orbitals (NM) are given 
in figure~\ref{fig3}. Due to a rather complicated structure and small
energetic resolution between the states, the LDOS is plotted in a
smaller energy window compared to the previous complexes, and
the lower HOMO orbitals (e.g.~$\pi$) are not shown, but their
charge density plots are given in figure~\ref{fig3}.

From the charge density plots for the NM state we conclude that the 
bonding $\delta$ and $\pi$ molecular orbitals can be distinctly
separated into three groups. The first group includes $\delta_0$ and
$\pi_2$ orbitals representing the bonding orbitals of separate VBz
(outside) and VBz$_{2}$ (inside). The second non-bonding $u$-group,
$\delta_u$ and $\pi_u$, is localized on the outer VBz and inner
VBz$_{2}$ parts of the molecule. The third $g$-type group is 
associated with the VBz$_{2}$ part of the molecule and is responsible
for the bonding in the molecule. Generally, the molecular orbitals
associated with the central V atom are lying lower in energy than those 
of the outer vanadii. 

For the NM case, the group of states close to the Fermi level
includes $\delta_0$, $d\sigma_1$ and $d\sigma_2$ orbitals. 
For the $d\sigma_{1}$ ($d_{z^2}$-type), mainly located on the central
V atom, the transition to the chain-like $z$-delocalized state can be
seen: the transfer of charge density from the inner V to the
outer ones is clearly visible. At the same time, the $d\sigma_{2}$
orbital is localized exclusively on the outer V-atoms. 

For the FM-case the occupied $d\sigma_{1}$ and $d\sigma_{2}$ contain
contributions from all three vanadii, which is not the case 
for the AFM1 configuration due to antiparallel spin-alignment.
The FM $d\sigma_{1}$ exchange splitting of 1.1~eV is smaller
than the corresponding value of 1.9~eV for $d\sigma_{2}$ level.
An exchange splitting of 1.9~eV for both $d\sigma$-states can
be observed for the AFM1 configuration. Concerning the 
$\delta$ states, the decrease in the exchange splitting can
be seen when increasing the hybridization from $\delta_0$ to 
$\delta_g$ in the FM case, while only the AFM1 $\delta_u$-state 
is exchange-split significantly. 

We have also considered a V$_4$Bz$_5$ complex, obtained by adding
an additional VBz molecule on top of the V$_3$Bz$_4$ compound.
The molecule is magnetic, and the ferromagnetic 
($\uparrow\uparrow\uparrow\uparrow$) solution is by
230~meV lower in energy than the nonmagnetic one.
We used the optimized ferromagnetic
atomic positions to calculate the total energy differences between
all possible collinear spin configurations. Our calculations
show that the high-spin FM solution with a total spin moment of
4$\mu_B$ is by 31~meV per V lower in energy, than the lowest of the
low-spin states.

\subsection{Magnetic anisotropy in V$_n$Bz$_m$ molecules}
\label{4.5}
The magnetism of V-Bz complexes and its 
development with increasing number of vanadii is an interesting
issue on its own. However, without an additional stabilizing
factor this magnetism is difficult to observe experimentally, considering
the scale of the energies involved, especially for smaller compounds.
The technological application of the molecules is impossible without 
the spin moment of the molecule being coupled to the lattice, and violation
of this will destroy such magnetic properties as half-metalicity or 
large magnetoresistance ratios~\cite{Maslyuk:06}.

The values of the MAE for the V$_{n}$Bz$_{m}$ compounds 
are presented in~table~\ref{table1}. While for the smaller
compounds the MAE is of the order of negligible 50~$\mu$eV per
V atom, it increases by an order of magnitude for $n>2$ 
accompanied by a change in the magnetization direction from $r$ to $z$ for 
large $n$. The value of 0.3~meV per V for an infinite VBz wire
is close to that of 0.1~meV, reported in Ref.~\cite{Xiang:06}. We conclude, 
that magnetism in V-Bz molecules of any length
can be stabilized only at very low temperatures, making their
technological application rather questionable without applying a strong magnetic
field from outside, despite their intriguing magnetism discussed before.

One way to increase the effective anisotropy barrier in V-Bz molecules
could be in depositing
the sandwiches on a substrate made of heavy elements, for instance, $5d$
transition-metals.
Such an approach was pioneered by Gambardella {\it et.~al.}
in Ref.~\cite{Gambardella:02}
for monoatomic Co chains deposited on a Pt surface, where a large MAE value
leads to stabilizing the long-range ferromagnetic order.
In this case the mechanism responsible for increase in the MAE
lies in a spin-polarization of the substrate atoms with a large spin-orbit 
interaction  due to
a large spin moment on the Co atoms~\cite{Baud:06}. 
For a VBz molecule, the spin moment of 1~$\mu_B$ 
of the V atom 
is effectively shielded from interacting with the
substrate by the outer benzene rings and a significant spin-polarization
of the substrate atoms is unlikely. In addition,
for spin-polarized transport applications~\cite{Maslyuk:06}
strong magnetism of the electrodes is required, which is not
the case for $4d$ and $5d$ transition-metal (TM) surfaces.  

A different route to achieve significant anisotropy energies is
to substitute the $3d$ vanadium atoms with heavier $4d$ or $5d$
transition-metals. Recently, it was reported, that in monoatomic
chains of $4d$ transition-metals giant values of MAE can be
observed~\cite{Mokrousov2006:4}. In monoatomic chains, however, the number of the nearest
neighbors for a magnetic atom is reduced to a minimal value of two,
resulting in enhanced magnetism with large spin moments. This is not
necessarily the case for Bz-based compounds, where a strong interaction
with large benzene molecules could, in principle, destroy the magnetism
of the transition-metals completely. In the next section we investigate
this idea.

\section{$\rm{Metal-Bz}$ molecules and $\rm{Metal-Bz}_{\infty}$ infinite
wires with $\rm{Metal=V,Nb,Ta}$}
\label{5}
In this section we study the magnetic properties of benzene and transition-metal (TM) 
half-sandwiches and infinite wires for the $4d$ and $5d$ TMs Nb and Ta, which are 
isoelectronic to V. 
This choice is based on the experimental
observation of Nb and Ta Bz-based compounds~\cite{Lyon2005:7.2,Rabilloud:03}, and the possibility
to tune the spin-orbit strength
in the molecules, without altering the electronic structure.
We have performed nonmagnetic and spin-polarized
calculations for the MBz half-sandwiches and considered for the
infinite wires a ferro- and antiferromagnetic ground state.

As seen in table~\ref{table2}, all three half-sandwiches are
magnetic with large magnetization energies $E_M$,
even for Ta with its extended $5d$ orbitals, and the smallest 
value of $E_M$ is obtained for the NbBz molecule. The
main features in the electronic configuration of the
molecules are preserved, going from V to Ta, with 
one unpaired electron occupying a non-bonding $d\sigma$
orbital, as explained in detail for the VBz sandwich in section~\ref{4.1}.
While the total spin moment constitutes a constant value
of 1~$\mu_B$ for all half-sandwiches (table~\ref{table2}),
its integrated value inside the muffin-tin sphere of the
metal atom shows a significant reduction with increasing
atomic number, from 1.09$\mu_B$ for V to 0.46$\mu_B$
for Ta, reflecting the increasing spread from $3d$ to $5d$ orbitals.  

\begin{figure}
\begin{center}
    \includegraphics[scale=0.55]{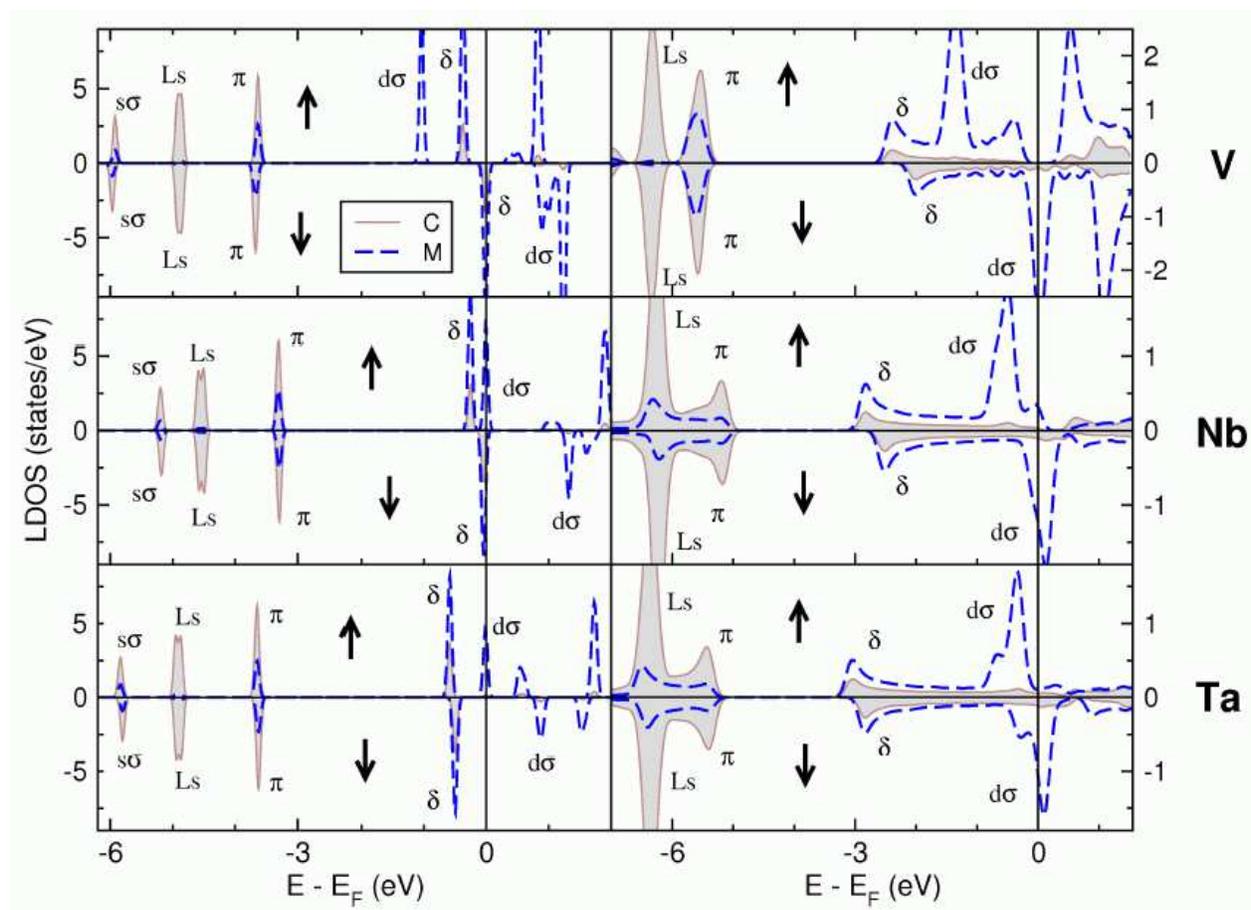}
\end{center}
\caption{Ferromagnetic LDOS for the half-sandwiches MBz (left panel) and infinite
wires MBz$_{\infty}$ (right panel) for M~=~V,~Nb and Ta.
Spin-up and spin-down channels are indicated by arrows.
The energetical positioning of the $\delta$ and $d\sigma$ states
strongly depends on the element. As compared to the DOS of the half-sandwiches,
broadening of the sharp molecular states into disperse bands for the infinite
wires can be seen.}
\label{fig7}
\end{figure}

Concerning the local densities of states (figure~\ref{fig7}),  
the energetic ordering of the $\delta$ and $d\sigma$ states, and
the exchange splitting of these states near the Fermi level
displays a significant dependency on the metal ion. 
Essential for applications are the states near the Fermi
level since they determine e.g.\ the spin-dependent transport
properties, and can be tuned by a particular  metal atom.
While the value of the $\delta$- and $d\sigma$-exchange splitting
is quite large for V, it decreases for Nb and Ta -- a consequence
of a larger spread of the corresponding $d$-orbitals.
The same phenomenon is responsible for the relative positioning
of the $d\sigma$ and $\delta$ states: a larger extent of the orbitals
leads to a larger overlap of the metal and benzene orbitals
with a stronger bonding, pushing the $\delta$ orbitals lower 
in energy with increasing atomic number.  

The properties of the infinite wires, which are the limiting case
of a very large number of the metal atoms in the molecules,
together with the information for the half-sandwiches, on the
other hand, provide an insight into the magnetism of complexes 
with increasing length.
As can be seen from table~\ref{table2}, the magnetization energies
and the spin moments differ significantly between MBz$_\infty$ wires
and MBz half-sandwiches. While for the
V wire the energy difference of 106~meV per metal ion in favor of magnetism
is of the order of that for the half-sandwich, it drops by an order 
of magnitude for Nb and Ta, constituting
small, but still sizable, values of 35.5 and 11.2~meV, respectively.   
This ensures us, that despite the large decrease in magnetization energy 
with increasing number of atoms in the Nb$_n$Bz$_m$ and Ta$_n$Bz$_m$
molecules, they will remain strongly magnetic even for very large $n$.
In the infinite wires of Nb and Ta the values of the total and 
muffin-tin spin moments are significantly reduced compared to the 
VBz$_\infty$ wire, reflecting a more delocalized nature
of the $d$-orbitals of these transition-metals.

The LDOS for the infinite wires (figure~\ref{fig7}) reflect
characteristic broadening of the sharp states in the DOS
for the half-sandwiches into bands. For example, the $\delta$
states of the single molecules correspond to strongly
dispersive $\delta$-bands with two-peak structure in the DOS characteristic 
for 1D-systems~\cite{Mokrousov2006:4}.
Among the two low-lying $Ls$ and $\pi$ 
groups of states the $Ls$-states show smaller dispersion,
as they originate from states pointing in-plane, rather
than along the chain axis. The $d\sigma$ states are also very
sharp, compared to $\delta$ states, reflecting their non-bonding
nature and small overlap along the $z$-axis. With increasing
nuclear number, the exchange splitting of the $\delta$ and
$d\sigma$ states decreases, in analogy to the single half-sandwiches,
the dispersion of the $\delta$ bands increases with their
left edge in the DOS moving to lower energies, while the 
$d\sigma$ band moves up in energy towards $E_F$.    
In contrast to finite molecules, which have a non-zero
HOMO-LUMO gap and are of 'insulating' character, the infinite M-Bz wires
are essentially metallic, with a non-zero density of states at the Fermi
level, originating mainly from a $d\sigma$-band in the spin-down channel.
The metallicity, more precisely, half-metallicity, of infinite V-Bz wire was
also theoretically predicted in Refs.~\cite{Maslyuk:06,Xiang:06}.

The importance of spin-orbit coupling in half-sandwiches molecules
and infinite wires with heavy $4d$ and
$5d$ TMs of Nb and Ta is reflected in the values of the 
orbital moments $\mu_L$ (table~\ref{table2}), calculated for 
two different magnetization directions.
For V compounds the absolute value of $\mu_L$
reaches 0.03$\mu_B$ for a half-sandwich, for the Nb
infinite wire it rises up to 0.07$\mu_B$,
and reaches as much as 0.15$\mu_B$ for the Ta half-sandwich. The negative
values of the orbital moments for most cases can be 
understood based on Hund's rules. The values of the MAE also
rise significantly when going from V to Ta, reaching a record
value of 7.5~meV for the TaBz compound, which is two orders
of magnitude larger than for VBz. For the infinite wires, the 
values of the anisotropy energy increase for V and Nb, while
for TaBz$_{\infty}$ the value of the 
MAE drops compared to the TaBz molecule, but the value of 3.2~meV
is still large. 
The easy axis of magnetization,
with an exception of NbBz$_{\infty}$, lies in the plane of the
benzenes for all the cases ($r$-case), and in general, follows the rule
that the direction of magnetization is that of the largest orbital 
moment, justified for $3d$~\cite{Bruno:89} and partly also applicable to 
$4d$ transition-metals~\cite{Mokrousov2006:4}.

We conclude, that substituting the vanadium ions
in MBz compounds by heavier Nb and Ta ions, which can be 
synthesized experimentally~\cite{Lyon2005:7.2,Rabilloud:03}, preserves the large magnetisation 
in these compounds with a pronounced tendency towards ferromagnetic
arrangement of the spins. In addition, it would allow to achieve much higher
values of the magnetic anisotropy energy, making it possible
to use these molecules for technological applications.

Another interesting effect of the spin-orbit coupling can be
seen for the NbBz$_{\infty}$ infinite wire, and can be 
related directly to the recently proposed transport phenomenon of the 
ballistic anisotropic magnetoresistance (BAMR)~\cite{Velev2005:7.4}. In contrast
to bulk materials, the mechanism of electronic transport changes
drastically when going to the nanoscale. When the dimensions
of the sample are smaller than the mean free path of the electrons,
the transport is ballistic, rather than diffusive, and the 
conductance shows a step-like behavior
with the applied voltage in units of $\frac{e}{h}$ for magnetic systems~\cite{Ono:99}.
In recent experiments performed on Ni ballistic
nanocontacts~\cite{Tsymbal:02}, a change of sign was found in the magnetoresistance
obtained for the field parallel and perpendicular to the current. 
The BAMR has recently been experimentally demonstrated for Co nanocontacts
in accordance to simple Co monowire band structure calculations~\cite{Tsymbal:07}.
The origin of this magnetoresistance anisotropy in the ballistic regime stems
from the fact that the conductance, e.g., in a perfect nanowire, depends on the number
of bands $N(E_F)$ crossing the Fermi energy $E_F$: 
$G = N(E_F)\cdot\frac{e^2}{h}$. The crucial point behind the BAMR effect is that the number of bands at $E_F$ 
can depend on the magnetization direction via the spin-orbit interaction,
which couples the orbital and spin moment, causing the projection of the
latter to be different depending on the direction of magnetization.

\begin{figure}
\begin{center}
    \includegraphics[width=11.5cm,keepaspectratio]{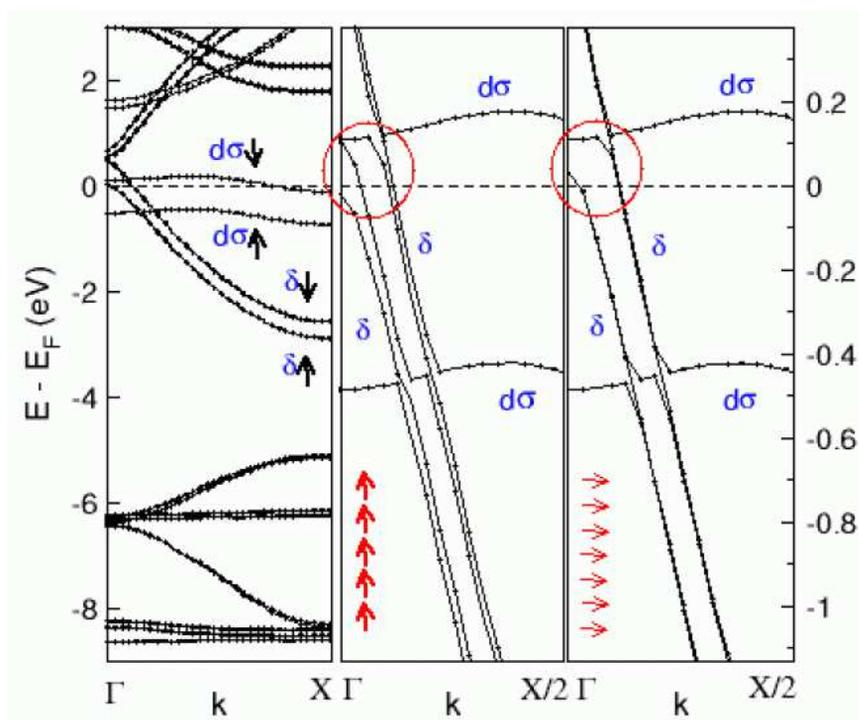}
\end{center}
\caption{ Calculated band structure of the infinite
Nb-Bz wire in the 1D Brillouin zone. Left panel --
band structure in the absence of the spin-orbit
coupling, with spin up and spin down indicated by
arrows; middle panel -- band structure with spin-orbit coupling
and the magnetization pointing along the $z$-axis;
right panel -- band structure with spin-orbit coupling 
and the magnetization in the planes of the benzenes.
Note, that the band structures in the middle and right
panels are given on a different energy scale, as compared
to the left panel, in the quarter of the 1D BZ.}
   \label{fig8}
 \end{figure}

It is important to realize, however, that the large desired values
of the anisotropic magnetoresistance are a consequence not only of
the electronic structure at the Fermi energy, but also of a large
spin-orbit splitting of the bands. For a NbBz infinite wire, e.g.,
a lucky combination of both occurs (figure~\ref{fig8}). Significant 
spin-orbit splitting of the bands at the Fermi energy leads to a different
number of states crossing the Fermi level for the two magnetization
directions. In particular, the edge of the $\delta$-band at the $\Gamma$-point 
of the Brillouin zone for the spin-up channel occurs exactly at the Fermi 
energy $E_F$, and undergoes a spin-orbit induced splitting when the 
direction of magnetization is along the $z$-axis, so that one of the SO-split bands
with the negative orbital character~\cite{Mokrousov2006:4}  gets completely occupied
and moves to lower energies. This is not the case for a magnetization
pointing in the plane of the benzenes, when no spin-orbit splitting of the 
$\delta$-bands can be seen, with two of the degenerate $\delta$-bands crossing
the Fermi energy. This allows to manipulate the conductance in the system by
an external magnetic field -- an example of ballistic anisotropic
magnetoresistance --  which  
opens a road for many possible spin-dependent transport applications.
In realistic systems, such as molecules sandwiched between two electrodes,
which rarely serve as ideal electron injectors, the conductance of the whole
system cannot be given anymore by simple counting of the bands at the Fermi
energy. The transport properties of the system will depend crucially on the
coupling to the leads as well as on subtle details of the contact geometry.
However, as follows from our calculations, for long Nb-Bz molecules and wires
the importance of spin-orbit coupling for states around the Fermi energy could
result in significant modulations of the overall conductance as a function of the
direction of the applied magnetic field, irrespective of the specific contact
geometry - an effect, which could be observed experimentally.

\section{Conclusions}

In our extended first-principles study we shed light on the electronic and magnetic
properties of metal-benzene sandwiches and infinite wires  
of V, Nb and Ta. We found that all the V$_{n}$Bz$_{m}$ ($n,m = 1,2,3,4$)
molecules and the infinite wire VBz$_{\infty}$ are
magnetic with a large gain in total energy compared to the
nonmagnetic solution. The magnetic ground state of all the molecules
is ferromagnetic with an exception of the V$_3$Bz$_4$ compound.
The competition of the super-exchange mechanism between the metal atoms via the
benzene molecules with the direct exchange between the metal ions results in small energy
differences between the ferromagnetic and antiferromagnetic solutions
for small $n$. The rigidity of the magnetism was investigated
with respect to the vibrations in the V$_2$Bz$_3$ compound.
With increasing number of vanadii in the complexes the
ferromagnetism becomes energetically more favorable, resulting in 
a monotonous increase of the total spin moment as a function of $n$, as observed
experimentally. The values of the spin moments and their qualitative behavior as a
function of the temperature is in good accordance with experiments as well.

We analyzed the stability of magnetism in V-based compounds by evaluating
a crucial quantity, the magnetic anisotropy energy (MAE), which fixes the
direction of the magnetization with respect to the lattice, making
the technological application of the sandwiches possible. We find
that for the molecules with $n<3$ the values of MAE are of the order
of tiny 50~$\mu$eV per metal atom, and rise with the number of vanadii
up to 0.3~meV in the infinite wire. We show that in order to increase the 
effective anisotropy barriers in the sandwiches it is desirable to use 
heavy Nb and Ta metal ions isoelectronic to V. Going from a single half sandwich to the
infinite wires the ferromagnetism in Nb-Bz and Ta-Bz compounds is preserved, while
the values of MAE increase by one to two orders of magnitude. Moreover, 
strong spin-orbit coupling in these molecules can result in intriguing 
physical effects with major consequences for spin-polarized transport and spintronics
applications. As an example, we demonstrate for a Nb-benzene infinite
wire the possibility of ballistic anisotropic magnetoresistance.

\ack

Two of us (N.A. and S.B.) acknowledge support by the DFG under the Priority 
   Programme 1137, Molecular Magnetism.
Financial support of the Stifterverband f\"ur die Deutsche
Wissenschaft and the Interdisciplinary Nanoscience Center
Hamburg is gratefully acknowledged.



\begin{table*}[t!]
\caption{Magnetism in V$_n$Bz$_m$ sandwiches.~$E_M$~($E_{AF}$)~stands 
for the total energy difference between the non-magnetic and 
ferromagnetic (ferromagnetic and antiferromagnetic) solutions.
Minus sign means that the magnetic (ferromagnetic) solution is lower in energy. The values of the spin moment $\mu_S$ are presented for the whole
molecules. MAE stands for the magnetic anisotropy energy
for the FM solution with the magnetization direction either perpendicular
to the wire axis ($r$) or along the wire axis ($z$). All values are given in
meV per V atom.}
\begin{tabular}{lrrrrrr}
\hline
   &{VBz}&{VBz$_{2}$}&{V$_{2}$Bz$_{3}$}&{V$_{3}$Bz$_{4}$}&{V$_{4}$Bz$_{5}$}&{VBz$_{\infty}$} \\
\hline
$E_M$ (meV/V)  & $-491$  & $-459$   &  $-415$  & $-300 $   & $-232$  &  $ -106 $      \\  
$E_{AF}$ (meV/V)& - &   -    &  $  -3$  & $  +5 $   & $ -31$  &  $  -57 $      \\  
$\mu_S$ ($\mu_B$) & $   1$   & $   1$   &  $   2$  & $   1 $   & $   4$  &  $  1_{\infty}$ \\ 
MAE (meV)                   & $0.05(r)$   & $0.05(r)$   &  $0.05(r)$  & $0.5(z)$   & $0.3(z)$  &  $0.3(r)$ \\ 
 \hline
\end{tabular}
\label{table1}
\end{table*}
\begin{table}
\caption{Magnetism in V, Nb and Ta half-sandwiches and infinite wires.
The values of the energies are given in meV per metal atom. The values
of the spin moments are given in $\mu_B$ per unit cell ($\mu_S$) and 
inside the muffin-tin of the metal atom ($\mu_S^{MT}$). The values
of the orbital moments are given in $\mu_B$ per metal atom for two different
directions of the magnetization ($\mu_L(z)$ and $\mu_L(r)$).
Minus sign of the orbital moment means that the direction of the orbital moment is opposite to that of the spin moment.}
\footnotesize{
   \begin{tabular}{c|ccccc|cccccc} \hline
\multicolumn{1}{c}{ } & \multicolumn{5}{c|}{MBz}        &\multicolumn{6}{|c}{MBz$_{\infty}$} \\ \hline \hline
                      & $E_M$ & $\mu_S$ ($\mu_S^{MT}$)  & $\mu_L(z)$ & $\mu_L(r)$ & MAE                 & $E_M$ & $E_{AF}$ &
$\mu_S$ ($\mu_S^{MT}$) & $\mu_L(z)$ & $\mu_L(r)$ & MAE   \\ \hline                                     
V                       & $-491$   & $1.0\,(1.09)$ & $0.00$       & $-0.03$      & $0.05(r)$ & $-110$ & $-57$ & $1.00\,(1.09)$ & $\,\,\,\,0.00$  & $-0.02$ & $0.3(r)$ \\ \hline
Nb                      & $-378$   & $1.0\,(0.70)$ & $0.00$       & $-0.05$      & $0.38(r)$ & $-35.5$ & $-21$ & $0.94\,(0.66)$ & $-0.07$ & $-0.02$ & $1.3(z)$ \\ \hline
Ta                      & $-388$   & $1.0\,(0.46)$ & $0.01$       & $-0.15$      & $7.50(r)$ & $-11.2$ & $-$  & $0.68\,(0.45)$ & $-0.05$ & $-0.08$ & $3.2(r)$ \\ \hline
   \end{tabular}
   }
   \label{table2}
\end{table}

\end{document}